\documentclass[single column,pre]{revtex4}
\usepackage{epsfig}
\usepackage{floatflt}
\usepackage{graphicx}
\usepackage{amsmath,amssymb,mathrsfs}
\usepackage{setspace}

\usepackage[T1]{fontenc}
\usepackage[latin1]{inputenc}
\usepackage[francais]{babel}

\usepackage{hyperref}
\begin{document}

\title{Unifying different interpretations of the nonlinear response in glass-forming liquids}

\author{P. Gadige$^{1,\star}$}
\author{S. Albert$^{1, \star \star}$}
\author{M. Michl$^2$}
\author{Th. Bauer$^{2, \star \star \star}$}
\author{P. Lunkenheimer$^2$}
\author{A. Loidl$^2$}
\author{R. Tourbot$^1$}
\author{C. Wiertel-Gasquet$^1$}
\author{G. Biroli$^{3,4}$}
\author{J.-P. Bouchaud$^5$}
\author{F. Ladieu$^{1 \dagger}$}
\email{$^\dagger$ Corresponding author. francois.ladieu@cea.fr ;$^\star$ Present address: Complex Fluids Lab, Soft Condensed Matter Group, Raman Research Institute, CV Raman Avenue, Sadashivanagar, Bangalore 560080, India; $^{\star \star}$ Present address: laboratoire de Physique, Ecole Normale Sup\'erieure de Lyon, 46 avenue d'Italie, 69364 Lyon Cedex 07, France; $^{\star \star \star}$ Present address: Institute for Machine Tools and Industrial Management, Technical University of Munich, 85748 Garching, Germany.}

\affiliation{$^1$ SPEC, CEA, CNRS, Universit\'e Paris-Saclay, CEA Saclay Bat 772, 91191 Gif-sur-Yvette Cedex, France.}
\affiliation{$^2$ Experimental Physics V, Center for Electronic Correlations and Magnetism, University of Augsburg, 86159 Augsburg, Germany.}
\affiliation{$^3$ IPhT, CEA, CNRS, Universit\'e Paris-Saclay, CEA Saclay Bat 774, 91191 Gif-sur-Yvette Cedex, France.}
\affiliation{$^4$ LPS, Ecole Normale Sup\'erieure, 24 rue Lhomond, 75231 Paris Cedex 05, France.}
\affiliation{$^5$ Capital Fund Management, 23 rue de l'Universit�, 75007 Paris, France.}

\date{\today}

\begin{abstract}
This work aims at reconsidering several interpretations coexisting in the recent literature concerning non-linear susceptibilities in supercooled liquids. We present experimental results on glycerol and propylene carbonate showing that the three independent cubic susceptibilities have very similar frequency and temperature dependences, both for their amplitudes and phases. This strongly suggests a unique physical mechanism responsible for the growth of these non-linear susceptibilities. We show that the framework proposed by two of us [BB, Phys. Rev. B \textbf{72}, 064204 (2005)], where the growth of non-linear susceptibilities is intimately related to the growth of ``glassy domains'', accounts for all the salient experimental features. We then review several complementary and/or alternative models, and show that the notion of cooperatively rearranging glassy domains is a key (implicit or explicit) ingredient to all of them. This paves the way for future experiments which should deepen our understanding of glasses.
\end{abstract}

\maketitle

\section{\label{part1} Introduction}

Glassy materials represent a very wide class of everyday materials ranging from molecular glasses to granular systems, and from polymers to colloids and foams. Yet the microscopic mechanisms leading to the spectacular increase of their relaxation time with temperature or density is still controversial. In particular, the existence of an underlying thermodynamic critical point, which would explain why {\it rigidity} develops in these systems, is a hotly debated issue \cite{BBRMP}. 

In the last fifteen years, however, some consensus has emerged about the existence of a growing length scale accompanying the slowing down of the dynamics of these various materials. Although anticipated by Adam \& Gibbs \cite{Ada65} more than $50$ years ago, the status of this length scale has remained elusive for a long time. For example, it was often argued that within the Mode-Coupling Theory of glasses the dynamical arrest phenomena are purely local \cite{Got92}. However, quite the contrary was shown 
in \cite{BBMCT,FPMCT,IMCT}. The Random First Order Transition (RFOT) theory provides a consistent framework to understand Adam \& Gibbs' intuition: a supercooled liquid should be thought of as a mosaic of locally rigid, but amorphous regions, the size of which increases as the temperature is reduced \cite{RFOT}. The necessity of a growing length scale in super-Arrhenius systems, an argument put forward by many,  was finally proved by Montanari \& Semerjian in \cite{MS}. These theoretical breakthroughs have spurred a flurry of experimental and numerical attempts to elicit this length scale, directly or indirectly -- see e.g. \cite{bookOUP}.

Among the different investigation tools, non-linear effects are especially interesting: the non-linear susceptibility is expected to have very different behavior when a genuine ``amorphous order'' sets in, as within RFOT, in contrast to the case of purely dynamical scenarii, such as provided by Kinetically Constrained Models (KCM) \cite{kcm}, for which non-trivial thermodynamic correlations are absent. In particular, based on an analogy with spin-glasses where the third order static susceptibility $\chi_3$ is known to diverge at the transition, two of us (BB) proposed in 2005 \cite{Bou05} that the non-linear a.c. susceptibility of glasses should peak at a frequency of the order of the inverse relaxation time, with a peak height that increases as the       
number $N_{\text{corr}}$ of molecules collectively involved in typical relaxation events. In the spirit of the fluctuation-response theorem, the increase of the peak of $\chi_3$ reveals the growth of quasi-static amorphous correlations in the system -see Eqs \ref{eq8},\ref{eq18} below-. 

The predictions of BB have been broadly confirmed using non-linear dielectric response in several experimental setups, first in glycerol \cite{Cra10}, then in several other glass formers \cite{Bau13} and in plastic crystals \cite{Mic16}, as well as for various pressures \cite{Cas15}. In all these studies, the temperature behavior of $N_{\text{corr}}$ (inferred from the peak of $\chi_3$) is in reasonable agreement with other, more indirect evidence \cite{Bau13,Mic16,Cas15,Ber05,Dal07}. These experiments have recently been extended to the fifth-order non-linear susceptibility $\chi_5$ in glycerol and propylene carbonate \cite{Alb16}, and are again fully consistent with the BB picture. In fact, the growth of the peak of $\chi_5$ as the temperature is reduced is stronger than that of $\chi_3$, and provides strong qualitative and quantitative evidence for the existence of an underlying critical point that drives the physics of supercooled liquids \cite{Alb16}. We also note that the non-linear mechanical response has also been studied in colloids \cite{Bra10,Sey16}, extending the BB results \cite{Bou05,Tar10} to that case as well.  

However, alternative theoretical interpretations have recently been proposed \cite{Ric16a,Ric16b}, invoking other effects to explain the non-linear effects, seemingly unrelated to the growth of $N_{\text{corr}}$. 
In order to clarify this issue, in the present paper we present additional experimental observations (Section \ref{part2}). We show that all three independent cubic a.c. susceptibilities have very similar frequency and temperature dependences, and their phases are related one to another. As we shall show (Section \ref{part3}), this 
is very natural if the physical origin is the same and due to the increase of $N_{\text{corr}}$, 
but it is instead at odds with simple phenomenological pictures. Furthermore, we show (Section \ref{part4}) that some of the alternative arguments can only explain the experimental results if some cooperative effects are present, as assumed by BB. 
    
\section{\label{part2} Three kinds of $\chi_3$ and their empirical behaviour}

\subsection{\label{part2-1} Setup and definitions}

We first recall the general formalism defining the third-order susceptibilities by introducing the time-dependent kernel $\chi_3$, relating polarization and electric field as follows: 
\begin{eqnarray}
\frac{P(t)-P_{\text{lin}}(t)}{\epsilon_0} &=&   
\iiint  \chi_{3}(t-t'_1, t-t'_2,t-t'_{3}) \times \nonumber \\
\  &\ & \times  E(t'_1) E(t'_2) E(t'_{3}) dt'_1 dt'_2 dt'_{3} + ...
\label{eq5}
\end{eqnarray} 
where higher order terms in the field are not written because they correspond to higher-order susceptibilities and where $\epsilon_0$ is the vaccuum  dielectric constant. 
Note that the threefold convolution product contained in Eq. \ref{eq5} is a simple generalisation of the standard onefold convolution product used to express the linear response. 
In purely ac experiments where the magnitude of the oscillating field $E_{\text{ac}}$ (of 
angular frequency $\omega = 2 \pi f$) is varied, two cubic responses arise, at frequencies $\omega$ and $3 \omega$. If a static field $E_{\text{st}}$ is superimposed on top of $E_{\text{ac}}$, new cubic responses arise, both for even and odd harmonics. By setting $\delta P  \equiv P(E_{\text{ac}},E_{\text{st}}) - P_{\text{lin}}(E_{\text{ac}},E_{\text{st}})$ and keeping only the odd harmonics, we get:  

\begin{eqnarray}
\frac{\delta P}{\epsilon_0} &=& \frac{3}{4} \vert \chi_{3}^{(1)}\vert E_{\text{ac}}^3\cos{(\omega t - \delta_{3}^{(1)})} + \nonumber \\
\ &\ & + \frac{1}{4} \vert \chi_{3}^{(3)}\vert E_{\text{ac}}^3\cos{(3 \omega t - \delta_{3}^{(3)})} + \nonumber \\
\ &\ & + 3 \vert \chi_{2,1}^{(1)}\vert E_{\text{st}}^2 E_{\text{ac}}\cos{(\omega t - \delta_{2,1}^{(1)})}
\label{eq19}
\end{eqnarray} 
where we have used the threefold Fourier transform of the kernel introduced in Eq. \ref{eq5} and defined:
\begin{eqnarray}
\vert \chi_{3}^{(1)}\vert  \exp{(-i \delta_{3}^{(1)})} &:=& \chi_3(-\omega, \omega,\omega),\nonumber \\
\vert \chi_{3}^{(3)}\vert  \exp{(-i \delta_{3}^{(3)})} &:=& \chi_3(\omega, \omega,\omega),\nonumber \\
\vert \chi_{2,1}^{(1)}\vert  \exp{(-i \delta_{2,1}^{(1)})} &:=& \chi_3(0,0,\omega).
\end{eqnarray}
For any cubic susceptibility -- generically noted $\chi_3$ -- the corresponding dimensionless cubic susceptibility $X_3$ is defined as :
\begin{equation}
X_3 = \frac{k_B T}{\epsilon_0 \Delta \chi_1^2 a^3} \chi_3
\label{eq8}
\end{equation}
where $\Delta \chi_1$ is the ``dielectric strength'', i.e. $\Delta \chi_1 = \chi_{\text{lin}}(0) - \chi_{\text{lin}}(\infty)$ where $\chi_{\text{lin}}(0)$ and  $\chi_{\text{lin}}(\infty)$ are respectively the linear susceptibility at zero and infinite frequency. Note that $X_3$ has the great advantage to be both dimensionless and independent of the field amplitude. Similar quantities can be defined for dimensionless fifth order responses, as explained in Ref. \cite{Alb16}.

\begin{figure}[t] 
\includegraphics[keepaspectratio,width=9.3cm]{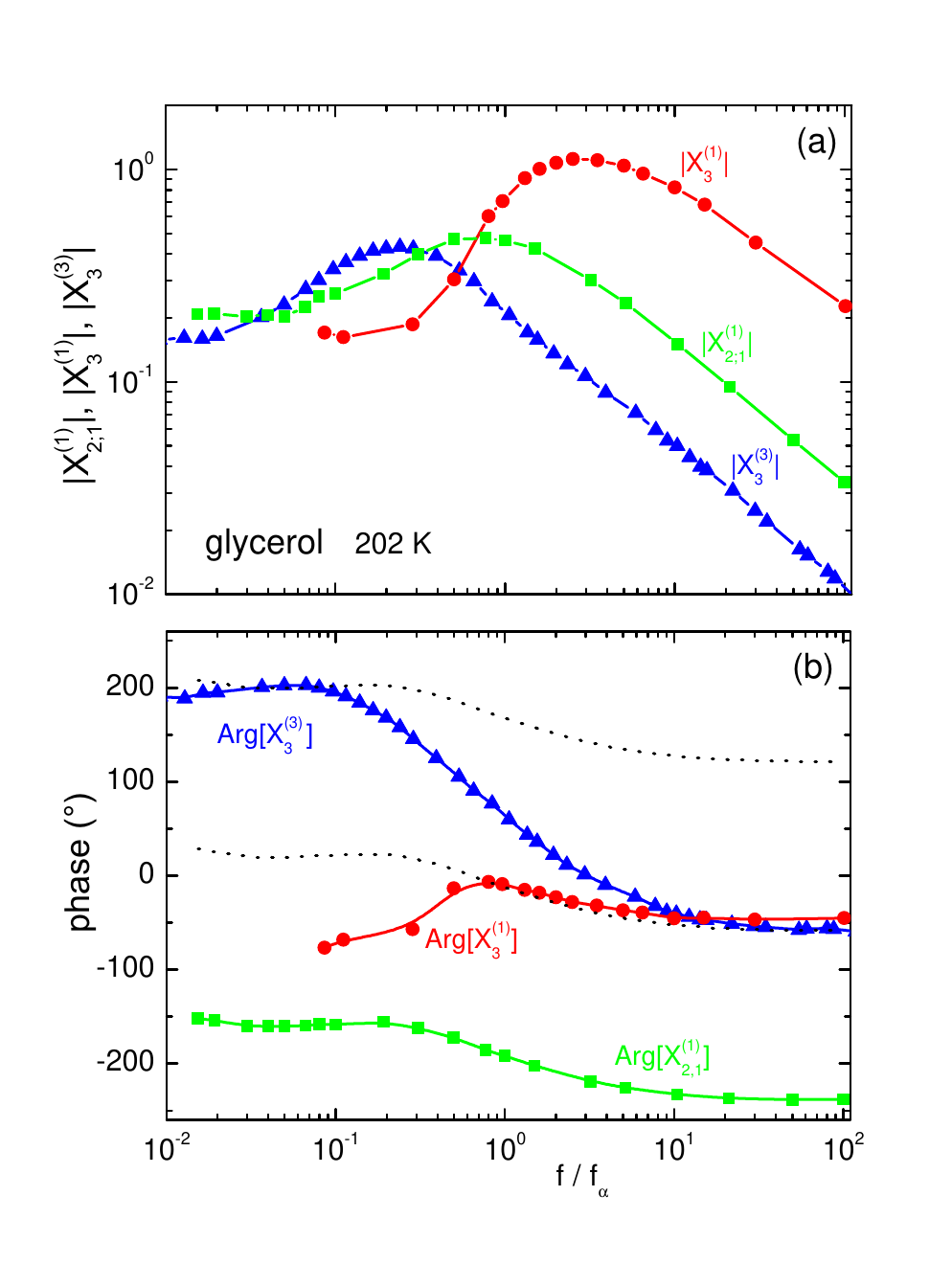}
\caption{(Color Online) Comparison of three cubic dimensionless susceptibilities of glycerol recorded at the same temperature with the same samples. The field amplitude is varied in $[2$MV/m$;5$MV/m$]$  (Saclay setup). Amplitudes (a) and phases (b) are shown -lines are just guide to the eyes-. $X_3^{(1)}$ and $X_3^{(3)}$ correspond to pure ac experiments, respectively to the first and third harmonics cubic response, $X_{2,1}^{(1)}$ is the cubic response at 
the first harmonics when a dc field is superimposed to an ac field. Note the strong similarities between these three quantities. Moreover the temperature behaviour of the peak is the same for the three dimensionless susceptibilities reported here. The two dotted lines in panel (b) correspond to a global shift of $\mathrm{Arg}[X_{2,1}^{(1)}]$ either by $180$ degrees -supporting Eq. \ref{eq20}- or by $360$ degrees -see text-.} 
\label{fig1}
\end{figure}

\begin{figure}[t] 
\includegraphics[keepaspectratio,width=9.6cm]{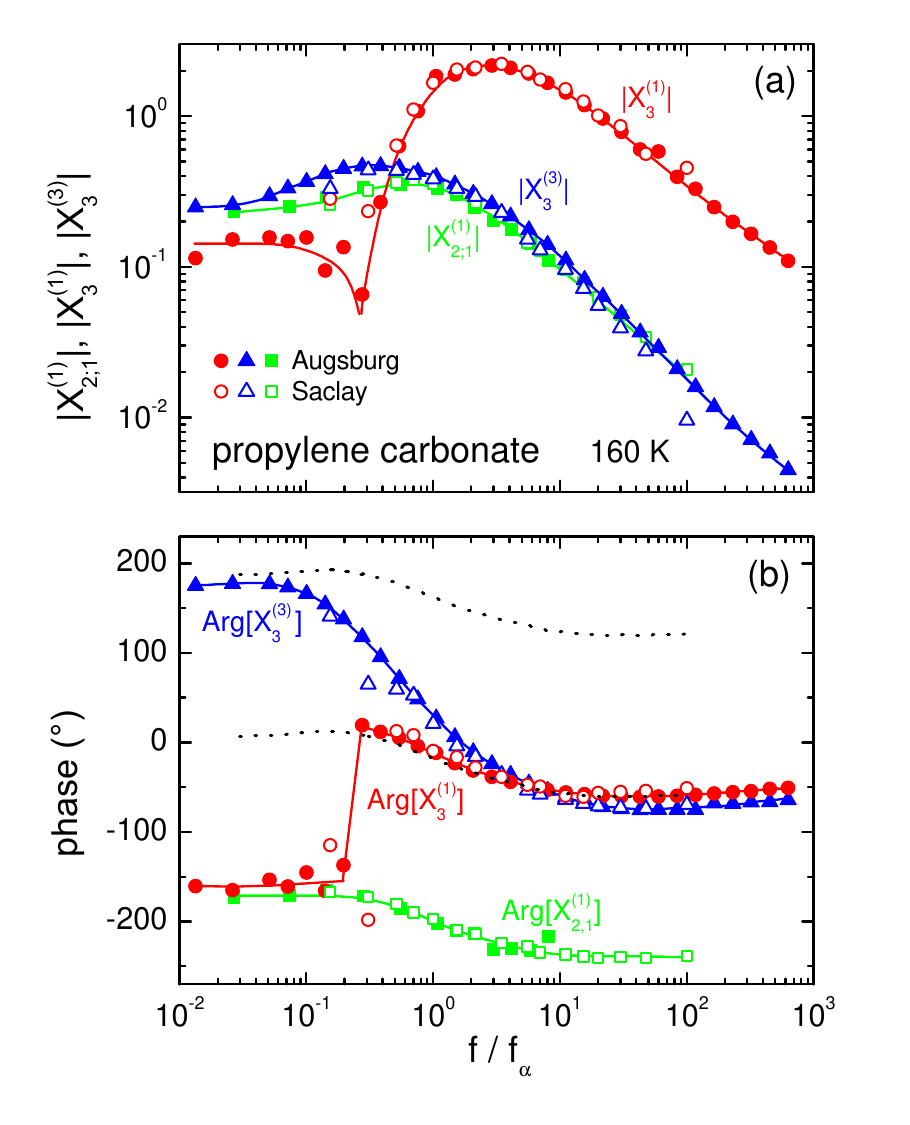}
\caption{(Color Online) Same as Fig.1 but for propylene carbonate -lines are just guide to the eyes-. The field amplitude is varied in $[2$MV/m$;5$MV/m$]$ in the Saclay setup and in $[2.6$MV/m$;3.5$MV/m$]$ in the Augsburg setup.} 
\label{fig2}
\end{figure}

Considering Eq. \ref{eq5}, one anticipates theoretically that the three cubic susceptibilities are closely related, since they all originate from the same pulse response function. However it was claimed in Refs \cite{Ric16a,Ric16b} that the several unrelated effects contributing to the three $X_3$'s could be singled out by a separate measurement of each cubic susceptibility. In this work we unveil the deep similarities existing between $X_3^{(1)}, X_3^{(3)}$ and $X_{2,1}^{(1)}$ that we have experimentally determined in glycerol and propylene carbonate -which are archetypical glass formers. 
The experiments were done in the Augsburg and Saclay setups described elsewhere \cite{Alb16,Bau13,Bru11}. For each data point of Figs \ref{fig1}-\ref{fig2}, the field was varied to ensure that the data obey Eq. \ref{eq19} -for the specific case of $X_{2;1}^{(1)}$ the ac field $E_{ac}$ was kept well below the static field $E_{st}$-. We briefly emphasize that the nonlinear effects reported here 
have been shown to be free of exogeneous effects: the global homogeneous heating of the samples by the dielectric energy dissipated by the application of the strong ac field $E_{\text{ac}}$ was shown to be fully negligible for $X_3^{(3)}$ as long as the inverse of the relaxation time $f_{\alpha}$ is $\le 1$kHz, see Ref. \cite{Bru10}. These homogeneous heatings effects were kept negligible in $X_3^{(1)}$ -- to which they contribute much more -- by keeping $f_{\alpha}$ below $10$Hz for the Saclay setup \cite{Bru11}, or by severely limiting the number $n$ of periods during which the electric field is applied (Augsburg setup, see \cite{Sup13}). The contribution of electrostriction was demonstrated to be safely negligible in Ref. \cite{Bru11}, both using theoretical estimates and
by showing that changing the geometry of spacers does not affect $X_3^{(3)}$. As for  the $\sim 0.5\%$ ionic impurities present in both liquids, we briefly explain that it has a negligible role, excepted at zero frequency where it might explain why the three $X_3$'s are not strictly equal, contrarily to what is expected on general grounds. Let us recall that on the one hand  it was shown that ion heating contribution  is fully negligible in $X_{2,1}^{(1)}$ (see the Appendix of Ref. \cite{Lho14}), on the other hand it is well known that ions affect the linear response $\chi_{\text{lin}}$ at very low frequencies (say $f/f_{\alpha} \lesssim 0.05$): this yields an upturn on the out-of-phase linear response $\chi^{\prime \prime}_{\text{lin}}$, which diverges as $1/\omega$ instead of vanishing as $\omega$ in an ideally pure liquid containing only molecular dipoles.  This is why we do not push our nonlinear measurements below $0.01 f_{\alpha}$, because at lower frequencies the nonlinear response is likely to be dominated by the ion contribution. In the same spirit, when measuring $X_{2;1}^{(1)}$, the static field was applied during a finite amount of time -longer than $1/\omega$- and its direction was systematically reversed to minimize any ionic migration effect. Finally, to avoid mixing the cubic response of molecular dipoles with that of ions, we have not measured the cubic response obtained just by using a pure static field. Therefore we do not reach the zero frequency limit where, on general grounds, one expects all the cubic susceptibilities to be equal. We think this is the reason why in Figs  \ref{fig1}-\ref{fig2} the three cubic susceptibilities are still slightly different even at the lowest frequencies that we have investigated.

\subsection{\label{part2-2} Experimental results}

Fig. \ref{fig1} shows the behaviour of the three cubic susceptibilities for supercooled glycerol at $202$K where the inverse relaxation time $f_{\alpha} \simeq 2$Hz. Fig.\ref{fig2} reports the same results for propylene carbonate at $160$K where $f_{\alpha} \simeq 0.2$Hz. The top graphs of Figs \ref{fig1}-\ref{fig2} display the modulii of the cubic suceptibilities while the bottoms graphs show the associated phases. We find four salient experimental features:
\begin{enumerate}
\item 
 The three moduli have a humped shape in frequency, with a peak located in the region of $f_{\alpha}$, namely at $0.22 f_{\alpha}$ for $X_3^{(3)}$, at $0.8 f_{\alpha}$ 
for $X_{2,1}^{(1)}$, and at $2.5 f_{\alpha}$ for $X_3^{(1)}$. These three numerical prefactors are only slightly different in propylene carbonate. Above the peak (higher frequencies), the modulii of the three cubic susceptibilities decrease as $\sim (f_{\alpha}/f)^{0.6}$ for glycerol and as $\sim (f_{\alpha}/f)^{0.7}$ for propylene carbonate. Below the 
peak (lower frequencies), the modulii fall down when decreasing frequency and become independent of frequency when $f/f_{\alpha} \le 0.05$. We refer to this low frequency domain as the ``plateau'' region \cite{note1}.

\item  The temperature dependence of the three dimensionless susceptibilities is significantly stronger around and above the hump than in the ``plateau'' region. Around and above the hump, the three cubic susceptibilities have a temperature dependence which is very close to that of $T \chi_T := T \vert \partial \ln{f_{\alpha}}/(\partial T)\vert$, see refs \cite{Cra10,Bru11,Bau13,Lho14,Cas15}.  Note that owing to the limited temperature interval accessible to experiments, one cannot distinguish clearly between $T \chi_T$ and $T^2 \chi_T$, see Refs. \cite{Bau13,Mic16,note6}. As for the ``plateau'' region, its temperature dependence is much weaker. It was convincingly shown in Refs. \cite{Cra10,Bru11,Lho14} that for glycerol, $X_{3}^{(3)}$ and $X_{2,1}^{(1)}$ do not depend on temperature in the plateau region, up to the experimental accuracy of $\pm 3\%$ per data point. This is also the case for propylene carbonate \cite{Bau13} where the plateau region lies in the same range of $f/f_{\alpha}$. Last but not least, the measurements of $X_3^{(3)}$ at various pressures was achieved in Ref. \cite{Cas15} and it was shown that the effect of pressure can be related to the effect of temperature. 
 
\item The phases of the three cubic responses basically do not depend explicitly on temperature \cite{Cra10,Bru11}, but only on $u=f/f_{\alpha}$, through
a master curve which depends only on the precise cubic susceptibility under consideration, see Figs. \ref{fig1}-\ref{fig2}. These master curves have the same qualitative shape as a function of $u$ in both glycerol and propylene carbonate. We note that the phases of the three cubic susceptibilities are related to one another. In the plateau region all the phases are equal (see the upper dotted line in Figs. \ref{fig1}-\ref{fig2}), which is easily understood because at low frequency the system responds adiabatically to the external field. At higher frequencies,  
we note that for glycerol (expressing the phases in radians):
\begin{eqnarray}
\mathrm{Arg} \left[ X_{3}^{(1)} \right] & \approx & \mathrm{Arg} \left[ X_{2,1}^{(1)} \right]+\pi \quad \mathrm{for} f/f_{\alpha} \ge 0.5;  \label{eq20} \\
\mathrm{Arg} \left[ X_{3}^{(1)} \right] & \approx & \mathrm{Arg} \left[ X_{3}^{(3)} \right] \qquad \mathrm{for} f/f_{\alpha} \ge 5  \label{eq21} 
\end{eqnarray}
which are quite non trivial relations, which holds also for propylene carbonate.

\item In the phase of $\chi_3^{(1)}$ of propylene carbonate (Fig. \ref{fig2}), a jump of $\pi$ is  observed which is accompanied by the indication of a spikelike minimum in the modulus -more details are given in the section \ref{partZ}-. A similar jump may also be present in glycerol (Fig. \ref{fig1}). We observe that this jump in the phase happens at the crossover between the T-independent ``plateau''  and the strongly T-dependent hump. More precisely in the ``plateau'' region one observes a reduction of the real part of the dielectric constant $\chi_{\text{lin}}'$, while around the hump $\chi_{\text{lin}}'$ is enhanced. At the frequency of the jump, both effects compensate and this coincides with a very low value of the imaginary part of $X_{3}^{(1)}$.
\end{enumerate}

Apart from this jump of $\pi$ which seems specific to $\chi_3^{(1)}$, the similarity between the three cubic susceptibilities reported in Figs. 1-2 puts strong constraints on the underlying physical mechanisms leading to an increase of the peak height with temperature.  

\section{Accounting for experimental results within the BB framework \label{part3}}

We now briefly explain why the aforementioned findings are in fact consistent with the theoretical framework put forward in BB \cite{Bou05}. The idea is that provided 
$f \gtrsim
 f_\alpha$, i.e. for processes faster than the relaxation time, one cannot distinguish between a truly frozen glass and a still flowing liquid. If some amorphous order is present in the glass phase, then non-trivial spatial correlations should be present and lead to anomalously high values of non-linear susceptibilities. If these spatial correlations extend far enough to be in a scaling regime, one expects the non-linear susceptibilities to be dominated by the glassy correlations and given by \cite{Bou05,Alb16}: 
\begin{equation}
X_{2k+1}^{\mathrm{glass}}(f,T) = [N_{\text{corr}}(T)]^k \times {\cal H}_k\left(\frac{f}{f_\alpha}\right)
\label{eq18}
\end{equation}
where the scaling functions ${\cal H}_k$ do not explicitly depend on temperature, but depends on the kind of susceptibility that is considered, i.e. $X_3^{(1)}$, $X_3^{(3)}$ 
or $X_{2,1}^{(1)}$ in the cubic case $k=1$. Since this ``glassy'' contribution has been shown to be the most divergent one \cite{Bou05,Tar10}, it should dominate over the other contributions to $X_{2k+1}$ as long as one does not enter in the low frequency regime $f \ll  f_{\alpha}$. In the latter regime, relaxation has happened everywhere in the system, 
destroying amorphous order \cite{note2} and the associated anomalous response to the external field and ${\cal H}_k(0)=0$. In other words, in this very low frequency regime, every molecule behaves independently of others and $X_{2k+1}$ is dominated by the ``trivial'' Langevin response of effectively independent molecules -- see \cite{note1} for a refined discussion. Due to the definition adopted in Eq. \ref{eq8}, the trivial contribution to $X_{2k+1}$ should not depend on temperature (or very weakly) . Hence, provided $N_{\text{corr}}$ increases upon cooling, there will be a regime where the glassy contribution $X_{2k+1}^{\mathrm{glass}}$ should exceed the trivial contribution, leading to humped-shape non-linear susceptibilities, peaking at $f_{\text{\text{peak}}} \sim f_\alpha$, where the scaling functions ${\cal H}_k$ reaches its maximum. Focusing on the three salient experimental facts discussed in the previous section, we find:
 \begin{enumerate}
\item Due to the fact that ${\cal H}_k$ does not depend explicitly on $T$, the value of $f_{\text{peak}}/f_{\alpha}$ should not depend on temperature, consistently with the experimental behavior. 
\item Because of the dominant role played by the glassy response for $f \ge f_{\text{peak}}$, the $T$-dependence of $X_{2k+1}$ will be much stronger above $f_{\text{peak}}$ than in the trivial low-frequency region. 
\item Last, because non-linear susceptibilities are expressed in terms of scaling functions, 
it is natural that the behaviour of their modulii and phases are quantatively related especially at high frequency 
where the "trivial" contribution can be neglected, consistently with Eqs. \ref{eq20}-\ref{eq21} (see below for a more quantitative argument in the context of the so-called ``Toy model'') \cite{note9}.  
\end{enumerate}

Let us again emphasize that the BB prediction relies on a scaling argument, where the correlation length $\ell$ of amorphously ordered domains is (much) larger than 
the molecular size $a$. This naturally explains the similarities of the cubic responses in microscopically very different liquids such as glycerol and propylene carbonate, as well as 
many other liquids \cite{Bau13,Cas15}. Indeed the microscopic differences are likely to be wiped out for large $\ell$, much like in usual phase transitions. 

Throughout this paper, we will not interpret $N_{\text{corr}}$ as a purely dynamical correlation volume, but as a static correlation volume, elicited by a ''quasi-static'' non-linear response (the frequency of the hump is indeed often lower than $f_{\alpha}$, see section \ref{part2}). This interpretation may seem surprising at first sight since theorems relating (in a strict sense) nonlinear responses to high-order correlations functions only exist in the static case, and therefore cannot be straightforwardly used to interpret the humped shape of $X_3$ (and of $X_5$) observed experimentally. 

This is why each theory of the glass transition must be inspected separately \cite{Alb16} to see whether or not it can account for the anomalous behavior of nonlinear responses observed in frequency and in temperature. The case of the family of KCM  is especially interesting since dynamical correlations, revealed by e.g. four-point correlation functions, exist even in the absence of a static correlation length. However in the KCM family, we do not expect any humped shape for nonlinear responses \cite{Alb16}. This is not the case for theories (such as RFOT or Frustration theories) where a non-trivial thermodynamic critical point drives the glass transition: in this case the incipient amorphous order allows to account \cite{Alb16} for the observed features of $X_3$ and $X_5$. This is why we think that in order for $X_3$ to grow some incipient amorphous order is needed, and we expect dynamical correlations in strongly supercooled liquids to be driven by static (``point-to-set'') correlations \cite{note12} --this statement will be reinforced by what we shall find in section \ref{part4-3}.

Notably, we find that the temperature dependence of $N_{\text{corr}}$ inferred from the height of the humps of the three $X_3$'s are compatible with one another, and closely related to the temperature dependence of $T \chi_T$, which was proposed in Refs. \cite{Ber05,Dal07} as a simplified estimator of $N_{\text{corr}}$ in supercooled liquids. The convergence of these different estimates, that rely on  general, model-free theoretical arguments, is a strong hint that the underlying physical phenomenon is indeed the growth of collective effects in glassy systems -- a conclusion that we shall reinforce by analyzing other approaches.

\section{\label{part4} Other phenomenological approaches}

A valid criticism of the general BB prediction is that the analytical expression of the scaling functions ${\cal H}_k$ is unknown, except for $k=1$ within MCT, where some analytical progress is possible \cite{Tar10}. In particular, only the $T$ dependence of $N_{\text{corr}}$ can be extracted from the experiments using Eq. \ref{eq18}, but not its absolute value. Moreover, since Eq. \ref{eq18} is only valid in the limit $N_{\text{corr}} \gg 1$, it may happen that other subleading contributions to $X_3^{\mathrm{glass}}$ are relevant in the limited range of temperatures available in practice. Some simplified, schematic models have therefore been proposed to compute explicitly the different cubic susceptibilities. 
We show that in each of them $N_{\text{corr}}$ is a key ingredient, either explicit, or implicit. For the sake of brevity, we concentrate on physical arguments, and postpone further analytical developments to the Appendix -we however recall some quantitative limitations in the first subsection. The first and third subsections are definitely phenomenological descriptions, whereas the second one starts from a solid thermodynamical relation recently pin-pointed by Johari, which, when coupled with the well known Adam-Gibbs relation, provides a physically motivated specification of the BB mechanism. 

\subsection{\label{part4-1} The Toy model and the Pragmatical model}

The ``Toy model'' has been proposed in Refs. \cite{Lad12,Lho14} as a simple incarnation of the BB mechanism, while the ``Pragmatical model'' is more recent \cite{Buc16a,Buc16b}. Both models start with the same assumptions: (i) each amorphously ordered domain containing $N_{\text{corr}}$ molecules has a dipole moment $\propto \sqrt{N_{\text{corr}}}$, leading to an anomalous contribution to the cubic response $X_{3}^{\mathrm{glass}} \propto N_{\text{corr}}$; (ii) there is a crossover at low frequencies towards a trivial cubic susceptibility 
contribution $X_{3}^{\mathrm{triv}}$ which does not depend on $N_{\text{corr}}$. We note \textit{en passant} that the ``Toy model'' predicts generally \cite{Lad12} an anomalous contribution $X_{2k+1}^{\mathrm{glass}} \propto [N_{\text{corr}}]^k$. This naturally accounts for the results of Ref. \cite{Alb16} where it was shown that $X_5^{\mathrm{glass}} \propto [X_3^{\mathrm{glass}}]^2$ in glycerol and propylene carbonate, consistently with the BB predictions summarized in Eq. \ref{eq18}. 

More precisely, in the ``Toy model'' each amorphously ordered domain is supposed to live in a simplified energy landscape, namely an asymmetric double well potential with a dimensionless 
assymetry $\delta$, favoring one well over the other. The most important difference between the Toy and the Pragmatical model come from the description of the low-frequency crossover, see Refs \cite{Lad12} and \cite{Buc16b} for more details.  

On top of $N_{\text{corr}}$ and $\delta$, the Toy model uses a third adjustable parameter, namely the frequency $f^*$ below which the trivial contribution becomes dominant. In Ref. \cite{Lad12}, \textit{both the modulus and the phase} of $X_3^{(1)}(\omega, T)$ and of $X_3^{(3)}(\omega, T)$ in glycerol were well fitted by using $f^* \simeq f_\alpha/7$, $\delta = 0.6$ and, for $T=204$K, $N_{\text{corr}} = 5$ for $X_3^{(3)}$ and $N_{\text{corr}} = 15$ for $X_3^{(1)}$. In Ref. \cite{Lho14}, the behavior of $X_{2,1}^{(1)}(\omega, T)$ in glycerol was further fitted with the same values of $\delta$ and of $f^*$ but with $N_{\text{corr}} = 10$ (at a slighly different temperature $T=202K$). Of course, 
the fact that a different value of $N_{\text{corr}}$ must be used for the three cubic susceptibilities reveals that the Toy model is oversimplified, as expected. 
However, keeping in mind that the precise value of $N_{\text{corr}}$ does not change the behaviour of the phases, we note that the 
fit of the three experimental phases is achieved \cite{Lad12,Lho14} by using the very same values of $f^*/f_{\alpha}$ and of $\delta$. This means 
that Eqs. (\ref{eq20}-\ref{eq21}) are well accounted for by the Toy model by choosing two free parameters.  
This is a quantitative illustration of how the BB general framework does indeed lead to strong relations between the various non-linear susceptibilities, such as those contained in Eqs. \ref{eq20}-\ref{eq21}.

Let us mention briefly the Asymetric Double Well Potential (ADWP) model \cite{Die12}, which is also about species living in a double well of asymmetry energy $\Delta$, excepted that two key assumptions of the Toy and Pragmatical models are not made: the value of $N_{corr}$ is not introduced, and the crossover to trivial cubic response is not enforced at low frequencies. As a result, the hump for $\vert X_3^{(3)} \vert$ is predicted \cite{Die12,Die17} only when the reduced asymmetry $\delta =\tanh(\Delta/(2k_BT)$ is close to a very specific value, namely $\delta_c = \sqrt{1/3}$, where $X_3$ vanishes at zero frequency due to the compensation of its several terms. However, at the fifth order \cite{Die17} this compensation happens for two values of $\delta$ very different from $\delta_c$: as a result the model cannot predict a hump happening both for the third and for the fifth order in the same parametric regime, contrarily to the experimental results of Ref. \cite{Alb16}. This very recent calculation of fifth order susceptibility \cite{Die17} reinforces the point of view of the Toy and Pragmatical models, which do predict a hump occurring at the same frequency and temperature due to their two key assumptions ($N_{corr}$ and crossover to trivial nonlinear responses at low frequencies). This can be understood qualitatively: because the Toy model predicts \cite{Lad12} an anomalous contribution $X_{2k+1}^{\mathrm{glass}} \propto [N_{\text{corr}}]^k$, provided that $N_{corr}$ is large enough, the magnitude of this contribution is much larger than that of the small trivial contribution $X_{2k+1}^{\mathrm{triv.}} \propto 1$, and the left side of the peak of $\vert X_{2k+1} \vert$ arises just because the Toy model enforces a crossover from the large anomalous response to the small trivial response at low frequencies $f \ll f_{\alpha}$. As for the right side of the peak, it comes from the fact that $\vert X_{2k+1} \vert \to 0$ when $f \gg f_{\alpha}$ for the simple reason that the supercooled liquid does not respond to the field at very large frequencies. 

 \subsection{\label{part4-3} Entropic effects}

We recall the argument given by Johari in \cite{Joh13,Joh16}. Suppose a static electric field $E_{\text{st}}$ is applied onto a dielectric material at temperature $T$. 
By using the general relations of thermodynamics, one finds that a variation of 
entropy $\left[ \delta S \right]_{E_{\text{st}}}$ follows, which for small $E_{\text{st}}$ is given by:
\begin{equation}
\left[\delta S \right]_{E_{\text{st}}} \approx \frac{1}{2} \epsilon_0 \frac{\partial \Delta \chi_1}{\partial T} E_{\text{st}}^2 a^3,
\label{eq1}
\end{equation}
where $a^3$ is the molecular volume. Eq. \ref{eq1} holds generically for any material. However, in the specific case of supercooled liquids close enough to their glass transition temperature $T_g$, a special 
relation exists between the molecular relaxation time $\tau_{\alpha}$ and the configurational contribution to the entropy $S_c$. 
This relation, first anticipated by Adam and Gibbs \cite{Ada65}, 
can be written as :  
\begin{equation}
\ln{\frac{\tau_{\alpha}(T)}{\tau_0}}= \frac{{\Delta_0}}{T S_c(T)} 
\label{eq2}
\end{equation}
where $\tau_{0}$ is a microscopic time, and $\Delta_0$ is an effective energy barrier for a molecule. The temperature dependence of $T S_c(T)$ quite well captures the 
temperature variation of $\ln(\tau_{\alpha})$, at least for a large class of supercooled liquids \cite{Ric98}. 

We now follow Johari \cite{Joh13,Joh16} and we assume that $\left[\delta S \right]_{E_{\text{st}}}$ is dominated by the dependence of $S_c$ with field, see Appendix \ref{partA1} for further discussion of this important physical assumption. Combining Eqs. \ref{eq1}-\ref{eq2}, we find that a static field $E_{\text{st}}$ produces a shift of $\ln(\tau_{\alpha}/\tau_0)$ given by:
\begin{equation}
\left[\delta \ln{\tau_{\alpha}} \right]_{E_{\text{st}}} = -\frac{{\Delta_0}}{T S_c^2} \left[ \delta S \right]_{E_{\text{st}}}
\label{eq3}
\end{equation} 
We show in Appendix \ref{partA1} that this entropic effect gives a contribution to $X_{2,1}^{(1)}$, which we call $J_{2,1}^{(1)}$ after Johari. Introducing $x=\omega \tau_{\alpha}$, we obtain: 
\begin{equation}
J_{2,1}^{(1)} = -\frac{{k_B \Delta_0}}{6S_c^2} \left[\frac{\partial \ln{(\Delta \chi_1)}}{\partial T}\right] \left[ \frac{ \partial \frac{\chi_{\text{lin}}}{\Delta \chi_1} }{\partial \ln{x}}\right] \propto \frac{1}{S_c^2}
\label{eq9}
\end{equation}
where $\chi_{\text{lin}}$ is the complex linear susceptibility. 

Eq. \ref{eq9} deserves two comments. Firstly $\vert J_{2,1}^{(1)} \vert$ has a humped shaped in frequency with a maximum in the region of $\omega \tau_{\alpha} \simeq 1$, because 
of the frequency dependence of the factor $\propto \partial \chi_{\text{lin}}/\partial \ln{x}$ in Eq. \ref{eq9}. Second, the temperature variation of $J_{2,1}^{(1)}$ is overwhelmingly dominated by that of $S_c^{-1}$ because $S_c \propto (T-T_K)$ with $T_K$ the Kauzman temperature. 

In fact, following Adam and Gibbs original formulation \cite{Ada65}, the dynamics of a supercooled liquid comes from ``cooperatively rearranging regions'' (CRR). Assuming that these regions are compact (see \cite{RFOT,Gilles}, and \cite{Alb16} for a recent discussion of this point), the spatial extension $\ell$ of the CRR is related to the number $N_{\text{corr}}$ of molecules as $N_{\text{corr}} = (\ell/a)^d$, where $d$ is the dimensionality of space. Within the Adam-Gibbs picture, $S_c \propto 1/N_{\text{corr}}$, leading to:
\begin{equation}
J_{2,1}^{(1)} \propto N_{\text{corr}}^q, 
\label{eq12}
\end{equation}
with $q=2$ i.e. a stronger divergence than predicted by BB, but a similar qualitative relation between non-linear effects and glassy correlations. Taking into account more general relationships between $S_c$  and $N_{\text{corr}}$ we find that the possible values of $q$ are bounded between $2/3$ and $2$, see Appendix \ref{partA2-1}.

However, the Adam-Gibbs picture has been reformulated more convincingly within the RFOT theory of glasses, see \cite{RFOT} and \cite{Bou04}. This leads to more constrained results (see Appendix \ref{partA2-2}):
\begin{equation}
J_{2,1}^{(1)} \propto N_{\text{corr}}^q, \qquad q = 1 + \frac{\Psi - \theta}{d}, 
\label{eq12bis}
\end{equation}
where $\Psi$ is the barrier exponent and $\theta$ the surface tension exponent (see Appendix \ref{partA2-2}). We note \textit{en passant} that formally, Adam-Gibbs corresponds to $\theta=0$ and $\Psi=d$. Comming back to RFOT, the exponents $\theta$ and $\Psi$ should obey, on general grounds, the following bounds:
\begin{equation}
\frac{d}{2} \leq \theta \leq d-1.
\end{equation}
The upper bound is natural for a surface-tension exponent, whereas the lower bound is obtained if one takes into account
the existence of self-induced disorder: if $\theta < d/2$ amorphous order would be destroyed by the disorder, which is what one expects below the lower critical dimension. 
Concerning $\Psi$, arguments based on the free-energy landscape give $\Psi\ge \theta$ \cite{Bou04}. 
However, it is possible that these do not hold for the dynamical rearrangements responsible for relaxation; indeed numerical results seem to favor $\psi< \theta$ \cite{cavagnapsi}.
 From these bounds, one concludes \cite{values} that -for $d=3$- $q$ lies in the range $[1/3,3/2]$, 
where $q=1$ corresponds to the ``recommended'' RFOT values, $\Psi=\theta=d/2$. Note that $q=1$ precisely corresponds to the BB prediction, in which case entropic effects are a physically motivated picture of BB's mechanism. For the lowest values of $q \simeq 1/3$, the 
Johari contribution is actually expected to be really subleading with respect to BB's contribution. Indeed, in the BB framework, the only way for $X_3$ to grow slower than $N_{\text{corr}}$ is that glassy domains are non-compact \cite{Alb16}, a possibility that is difficult to accomodate both with RFOT and with the experimental results of Ref. \cite{Alb16}.

To summarize this subsection, the two key assumptions for computing Johari's entropy effect are that the field induced entropy variation mainly goes into the configurational part of the entropy, and that its effect can be evaluated by using the Adam Gibbs relation. We have found that the entropy contribution to $X_{2;1}^{(1)}$, called $J_{2;1}^{(1)}$, is similar to the general BB prediction both because of its humped shape in frequency -see Eq. \ref{eq9}-, and because it is directly related to $N_{corr}$ -see Eqs. \ref{eq12}-\ref{eq12bis}-. Additionaly, because $S_c$ is a static quantity, Eqs. \ref{eq12}-\ref{eq12bis} support our interpretation that $X_3$ is related to static amorphous correlations, as stated in the end of section \ref{part3}.

Let us add two remarks. One is about the extension of the above calculation to $X_3^{(1)}$ and to $X_3^{(3)}$. Such an extension is a key ingredient of the phenomenological model elaborated in Refs. \cite{Ric16a,Ric16b} and gives the main term allowing to fit the modulus of $X_3^{(3)}$ for glycerol \cite{Ric16b}. This extension came after several works \cite{Sam15,You15,Rie15,Sam16b} where the entropic effects were found to be consistent with the measured $X_{2;1}^{(1)}$ in various systems. Note that to perform this extension one needs to introduce a time dependent configurational entropy, which is nevertheless acceptable in the region $\omega \tau_{\alpha} \lesssim 1$, where the model is used. The second remark is that there must be other contributions to $X_{2,1}^{(1)}$ coming from, e.g. the field dependence of the energy barrier $\Delta_0$ or of the surface energy cost $\Upsilon$ in RFOT. Following the calculations in Appendix \ref{partA2-2}, this leads to a contribution that behaves as $N_{\text{corr}}^{\Psi/d}$, which is subdominant compared to $J_{2,1}^{(1)}$ as given by Eq. (\ref{eq12bis}). 
This illustrates that between the leading BB's contribution to $X_{2;1}^{(1)}$ and the mere trivial contribution, there is room for intermediate terms scaling more slowly than $N_{corr}$.

\subsection{\label{part4-2} The Box model}

The ``Box model'' is historically the first model of nonlinear response in supercooled liquids, designed to account for the Nonresonant Hole Burning experiments \cite{Sch96}. It 
assumes \cite{Sch96,Ric06,Wei07,Wan07,Kha11} that some heterogeneous heating happens within each of the amorphously ordered domain, and that the thermal time during 
which the dissipated heat is kept within each domain is as long as the dielectric relaxation time of the domain, i.e.  as long as \textit{seconds} close to $T_g$. According to Ref. \cite{Buc16b}, this seems to contradict physical intuition, since the size of amorphously ordered domains is only a few nanometers \cite{note10}. The Box model has been shown to give good fits of the imaginary part of $X_3^{(1)}$ for $f > f_{\alpha}$ in many glass forming liquids, see e.g. \cite{Ric06,Wei07,Wan07,Kha11}. It was shown also \cite{Bru11b} that the Box model is \textit{not} able to 
fit quantitatively the measured $X_3^{(3)}$ (even though some qualitative features are accounted for), and that the Box model only provides a vanishing contribution to $X_{2,1}^{(1)}$  \cite{Lho14}.

In recent works \cite{Ric16a,Ric16b}, the three experimental cubic susceptibilities have been argued to result from  a superposition of an entropic contribution and of a contribution coming from the Box model, plus a trivial contribution playing a minor role around the peaks of the cubic susceptibilities. More precisely, the hump of
$\vert X_{2,1}^{(1]} \vert$ and of $\vert X_3^{(3)} \vert$ would be mainly due to the entropy effect, contrarily to the hump of $\vert X_{3}^{(1)} \vert$ which would be due to the Box 
 model contribution. This means that very different physical mechanisms would conspire to give contributions of the same order of magnitude, with furthermore phases that have no reason to match as they do empirically, see Eqs. \ref{eq20}-\ref{eq21}: why should $X_3^{(1)}$ and $X_3^{(3)}$ have the same phase at high frequencies if their physical origin is different?
  
We see no reason for such a similarity if the growth of $X_3^{(1)}$ and $X_3^{(3)}$ are due to independent mechanisms. Having just related entropic effects to the increase of 
$N_{\text{corr}}$, everything becomes instead very natural if the Box model is recasted in a framework where $X_3^{(1)}$ is related to the glassy correlation volume. To do a first step in this direction, we show in Appendix \ref{partA3} that the Box model prediction for $X_{3}^{(1)}$ at high frequencies is proportional to the above Toy model prediction, 
 provided  $N_{\text{corr}}$ and $T \chi_T$ are proportional -- which is a reasonable assumption as explained in the end of section \ref{part3} and Refs. \cite{Bru11,Ber05,Dal07}.  In all, the only reasonable way to account for the similarity of all three cubic susceptibilities, demonstrated experimentally in Section \ref{part2}, is to invoke a 
common physical mechanism. As all the other existing approaches previously reviewed in this paper relate cubic responses to the growth of the glassy correlation volume, reformulating the Box model along the same line is, in our view, a necessity. 

\section{\label{part5}Conclusion}

In summary, we have compared three different cubic susceptibilities, in two different liquids, and found that they all behave very similarly in frequency and in 
temperature, both for their modulii and for their phases. This suggests a unique underlying physical mechanism, which we argue is the growth of a glassy correlation length, measuring the size of the domains where amorphous order sets in. The theoretical framework proposed by two of us \cite{Bou05} (BB) provides a consistent description for all cubic (and higher-order \cite{Alb16})
susceptibilities $X_3$. We have reviewed various phenomenological models that attempt to give a quantitative description of $X_3$. Although some of them are at first sight not compatible with the previous scenario 
and lead to puzzling physical predictions compared to our experiments, we explained 
why they are not in contradiction with the BB predictions. Excepted for the Box model where the task is not fully achieved, all the models can be actually recasted in such a way that the number of correlated molecules $N_{\text{corr}}$ appears (implicitly or explicitly) as a key ingredient. 

Having unified various approaches of nonlinear responses close to $T_g$, our work opens at least two new routes of research. Firstly, 
it would be very interesting to access $\chi_3$ (and $\chi_5$) in molecular liquids at higher temperatures, closer to the MCT transition, and/or for frequencies close to the fast $\beta$ process 
where more complex, fractal structures with $d_f<d$ may be anticipated \cite{IMCT,WolynesSchmalian}. Note that even though $X_{3}^{(1)\ \mathrm{or}\ (3)}$ are plagued by heating issues when  $f_{\alpha}$ is large, this is not the case for $X_{2,1}^{(1)}$ because a d.c. field yields negligible dissipation. As we have shown that the three cubic suceptibilities are driven by the same physics, it would be wise to choose $X_{2,1}^{(1)}$ to investigate the behavior of molecular liquids at high temperatures. Secondly, we could revisit the vast field of polymers by monitoring their nonlinear responses, which should shed new light onto the temperature evolution of the correlations in these systems. Therefore we think that there is much room to deepen our understanding of the glass transition by carrying out new experiments about nonlinear susceptibilities. 

\vskip 0.5cm

{\bf{ACKNOWLEDGEMENTS}}
We thank C. Alba-Simionesco, U. Buchenau, A. Coniglio, P.-M. D\'ejardin, R. Richert, G. Tarjus, and M. Tarzia for interesting discussions. This work in Saclay has been supported in part by ERC grant NPRGLASS, by a grant from the Simons Foundation (\#454935, Giulio Biroli), by the Labex RTRA grant Aricover and by the Institut des Syst\`emes Complexes ISC-PIF. This work in Augsburg was supported by the Deutsche Forschungsgemeinschaft via Research Unit FOR1394.

\section{Appendix \label{partA}}
\subsection{The spike-like minimum in $\vert X_3^{(1)} \vert$ for propylene carbonate \label{partZ}}

\begin{figure}[t] 
\includegraphics[keepaspectratio,width=9.3cm]{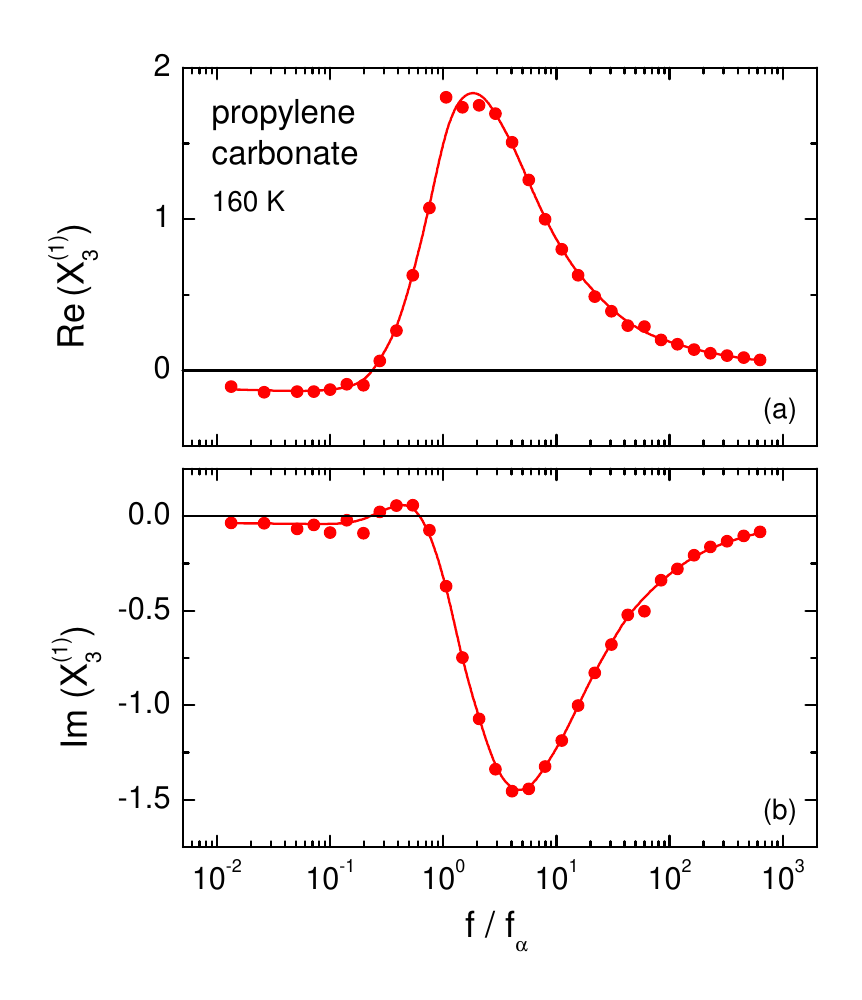}
\caption{(Color Online) Real and imaginary part of the $X_3^{(1)}$ data of propylene carbonate shown in Fig. 2. Lines are guides to the eyes.} 
\label{fig3}
\end{figure}

The spike-like minimum as indicated by the red line in Fig. 2(a) seems somewhat speculative. However, when plotting the same data set as real and imaginary part (see Fig. \ref{fig3}), it becomes obvious that $\vert X_3^{(1)} \vert$ becomes zero at a certain frequency, thus generating a negative spike in the logarithmic plot of Fig. 2-a: When the trivial response starts to dominate at low frequencies, the real part of $X_3^{(1)}$ must become negative because dielectric saturation causes a decrease of the dielectric constant $\chi'_{\text{lin}}$ at high fields (i.e., negative $\text{Re}(X_3^{(1)})$) instead of the increase seen at higher frequencies (positive $\text{Re}(X_3^{(1)})$). This causes $\text{Re}(X_3^{(1)})$ to cross the zero line. The imaginary part of $X_3^{(1)}$ also is close to zero in this region and, thus, the modulus $\vert X_3^{(1)} \vert$ also becomes extremely small at this crossover frequency.

\subsection{\label{partA1} Entropic effects}

The argument of Johari is to decompose the total entropy $S_{tot}$ of a supercooled liquid in its vibrational part $S_{vib}$ and its configurational part $S_c$. Then, because of the smallness of electrostriction effects in general \cite{Joh13,Joh16}, Johari deduces that the field induced variation of $S_ {vib}$ is much smaller than that of $S_c$. We note that this argument can be reinforced by an alternative reasoning: $S_{tot}$ can also be decomposed as the entropy of the crystal $S_{cryst}$ plus an excess entropy $S_{exc}$, where $S_{exc}$ contains the configurational entropy $S_{c}= f_S S_{exc}$ -the factor $f_S<1$ does not depend on the temperature \cite{Sam16} and will be disregarded hereafter. As for some archetypical glass formers -such as glycerol and propylene carbonate studied in the experimental section- the static value of the dielectric linear susceptibility $\Delta \chi_1$ is much larger in the supercooled liquid than in the crystal, it is very likely that $\partial \Delta \chi_1/{\partial T}$ is also much larger in the supercooled liquid than in the crystal. With the help of Eq. \ref{eq1}, this means that the field induced variation of $S_{cryst}$ is much smaller than that of $S_{exc}$ -and thus also of $S_{c}$. Of course the validity of this alternative argument is restricted to the subclass of glass formers where $\Delta \chi_1 \gg 1$ in the supercooled liquid state.

We now focus specifically onto the case where an a.c. field $E_{\text{ac}}$, of angular frequency $\omega$, is applied on top of the static field $E_{\text{st}}$. Consistently with the assumption of 
linear dielectric response allowing to derive Eq. \ref{eq1}, we express the polarization $P$ when $E_{\text{st}}=0$ as  $P(E_{\text{ac}},0,T) = \epsilon_0 \chi_{\text{lin}} E_{\text{ac}}$ where $\chi_{\text{lin}}$ is a complex 
quantity which, once scaled by $\Delta \chi_1$, depends mainly on $x = \omega \tau_{\alpha}$. Considering Eq. \ref{eq3}, we obtain:

\begin{equation}
P(E_{\text{ac}}, E_{\text{st}}, T) = P(E_{\text{ac}}, 0, T) + \frac{\partial P}{\partial \ln{\tau_{\alpha}}} \left[ \delta \ln{\tau_{\alpha}} \right]_{E_{\text{st}}} +...
\label{eq4}
\end{equation}
where higher order terms in $\delta \ln{\tau_{\alpha}}$ have been neglected. Because $P(E_{\text{ac}},0,T) \propto E_{\text{ac}}$ and $\delta \ln{\tau_{\alpha}} \propto E_{\text{st}}^2$, defining $\delta P_{E_{\text{st}}}  = P(E_{\text{ac}}, E_{\text{st}}, T) - P(E_{\text{ac}}, 0, T)$, one finds $\delta P_{E_{\text{st}}} \propto E_{\text{st}}^2E_{\text{ac}}$. This shows that $\delta P_{E_{\text{st}}}$ is cubic in the electric field.


Inserting Eq. \ref{eq1} into Eq. \ref{eq3}, and using Eq. \ref{eq4}, we find that the entropy variation induced by the static electric field yields a term $\propto E_{\text{st}}^2 E_{\text{ac}}$, 
i.e. it yields a contribution to $X_{2,1}^{(1)}$ because of the definition given in Eq. \ref{eq19}. 
We obtain for this entropic contribution to $\chi_{2;1}^{(1)}$: 

\begin{equation}
\chi_{2,1}^{(1),\delta S} = \left[ \frac{-1}{6} \right] \epsilon_0 \left[ \frac{\partial \Delta \chi_1}{\partial T} \right] a^3 \frac{{\Delta_0}}{S_c^2 T} \left[ \frac{\partial \chi_{\text{lin}}}{\partial \ln{x}} \right]
\label{eq7}
\end{equation}
where we remind our notation $x \equiv \omega \tau_{\alpha}$.

By using Eq. \ref{eq7} we find the entropy contribution $J_{2,1}^{(1)}$ to the dimensionless cubic susceptibility $X_{2,1}^{(1)}$. We get:

\begin{equation}
J_{2,1}^{(1)} = -\frac{{k_B \Delta_0}}{6S_c^2} \left[\frac{\partial \ln{(\Delta \chi_1)}}{\partial T}\right] \left[ \frac{ \partial \frac{\chi_{\text{lin}}}{\Delta \chi_1} }{\partial \ln{x}}\right] \propto \frac{1}{S_c^2}
\label{eq9b}
\end{equation}
where, as explained above, $\chi_{\text{lin}}$ is a complex quantity. 

As explained in the main text, Eq. \ref{eq9b} implies that $\vert J_{2;1}^{(1)} \vert$ is peaked for a frequency close to $f_{\alpha}$. We note that in an ideal 
gas of dipoles $\vert X_{2;1}^{(1)} \vert$ has \textit{no peak} at any frequency -see the works quoted in note \cite{note1}-. This comes from the fact that in an ideal gas of dipoles the relaxation time is insensitive 
to the static field, i.e. the last term of Eq. \ref{eq4} has to be neglected when computing $X_{2;1}^{(1)}$. Assuming a non zero 
value of $\left[ \delta \ln{\tau_{\alpha}} \right]_{E_{\text{st}}}$ is thus a highly non trivial assumption which calls for an explanation. We have argued, following Johari, 
that $\left[ \delta \ln{\tau_{\alpha}} \right]_{E_{\text{st}}}$ comes from entropic effects. In the case where this could be disputed -e.g., when $\Delta \chi_1 \simeq 1$ and/or when the factor $f_S$ turns out to be very far from $1$- we show briefly that a non zero value of $\left[ \delta \ln{\tau_{\alpha}} \right]_{E_{\text{st}}}$ is related to the glassy correlation volume. Indeed, one has:

\begin{equation}
\left[ \delta \ln{\tau_{\alpha}} \right]_{E_{\text{st}}} =  \left[ \frac{ \partial \ln{\tau_{\alpha}}}{\partial E_{st}^2}\right]_{T} E_{st}^2 
= - \left[ \frac{ \partial \ln{\tau_{\alpha}}}{\partial T}\right]_{E_{st}=0} {\cal{M}} E_{st}^2
\label{eq9d}
\end{equation}
where ${\cal{M}} = \partial T_g /\partial E_{st}^2$ expresses the shift of the glass transition temperature $T_g$ induced by $E_{st}$, i.e. the fact that the dielectric spectrum is uniformly shifted in frequency by the static field -- the minus sign in the last equality of Eq. \ref{eq9d} comes from the mapping between $P(T, E_{st})$ and $P(T-{\cal M}E_{st}^2, 0)$, see \cite{Lho14}. As a result:

\begin{equation}
\left[ \frac{ \partial \ln{\tau_{\alpha}}}{\partial E_{st}^2}\right]_{T}
= -{\cal{M}} \left[ \frac{ \partial \ln{\tau_{\alpha}}}{\partial T}\right]_{E_{st}=0} 
 = \frac{{\cal{M}}}{T} \vert T\chi_T \vert
\label{eq9e}
\end{equation}
which establishes that a non zero value of $\left[ \delta \ln{\tau_{\alpha}} \right]_{E_{\text{st}}}$ must be related to $T\chi_T$, i.e. to the glassy correlation volume, as advocated in Refs. \cite{Ber05,Dal07}. Having briefly evoked the case where the origin of the non zero   
value of $\left[ \delta \ln{\tau_{\alpha}} \right]_{E_{\text{st}}}$ is questionnable, we now come back to the case where this origin is the entropy effect pin-pointed by Johari.

\subsection{\label{partA2} Relations between configurational entropy and length scales}
\subsubsection{\label{partA2-1} The Adam-Gibbs case}

When lowering $T$, a supercooled liquid becomes increasingly viscous, and its dynamics comes from ``cooperatively rearranging regions'', to quote the original expression of 
Adam and Gibbs \cite{Ada65}. Expanding on \cite{Alb16}, as well as on several theoretical approaches \cite{RFOT,Gilles}, we shall assume that these regions are compact, i.e. their 
spatial extension $\ell$ is related to their number $N_{\text{corr}}$ of molecules by $N_{\text{corr}} = (\ell /a)^d$, where $d$ is the dimensionality of space.

Coming back to the original argument of Adam Gibbs \cite{Ada65} readily gives a \textit{lower} bound for $N_{\text{corr}}$. Indeed, owing to its extensive character, 
the configurational entropy of a domain of size $\ell$ is $S_c (\ell /a )^d$. For this domain to be able to relax, at least two states must be available, and thus the 
aforementioned configurational entropy cannot be smaller than  $k_B \ln{2}$. As a result:

\begin{equation}
\left(\frac{\ell}{a}\right)^d \ge \frac{k_B \ln{2}}{S_c} \quad ; \quad  \frac{k_B \ln{2}}{S_c} \le N_{\text{corr}} 
\label{eq10}
\end{equation}

Besides, by using Refs. \cite{Bou04,New02,Fis03}, one can find an \textit{upper} bound for $N_{\text{corr}}$. 
For a given domain of size $\ell$ where $N_{\text{corr}}$ molecules relax cooperatively, the argument comes from 
the comparison of the number of accessible states $\mu S_c(\ell/a)^d$ with the number of different boundary conditions $\lambda (\ell/a)^{d-1}$ -- here $\mu$ and $\lambda$ are constants--. 
The latter must be larger than the former, otherwise there are not enough boundary conditions to select each of the accessible states. This would mean that when freezing all the 
 molecules of the system excepted those inside the domain with size $\ell$, one cannot define a prefered state, since many states are possible. This contradicts the assumption that within the considered domain all the molecules are in a well defined state. To avoid this contradiction, one needs to write: 

\begin{equation}
\lambda \left(\frac{\ell}{a}\right)^{d-1} \ge \mu S_c \left( \frac{\ell}{a} \right)^d \quad ; \quad N_{\text{corr}} \le \left[ \frac{\lambda}{\mu S_c} \right]^d
\label{eq11}
\end{equation}

By using the two boundaries obtained in Eqs. \ref{eq10}-\ref{eq11}, we get with $d=3$: 

\begin{equation}
\frac{1}{S_c^2} = \Gamma \left[ N_{\text{corr}} \right]^q \hbox{\ with\ } \frac{2}{3} \le q \le 2 
\label{eq12b}
\end{equation}
where both the proportionality constant $\Gamma$ and the exponent $q$ should not depend on temperature owing to the fact that $S_c$ and $N_{\text{corr}}$ are the only temperature dependent quantities 
in the aforementioned inequalities. 

We now combine Eq. \ref{eq12b} and Eq. \ref{eq9b} to obtain: 

\begin{equation}
J_{2,1}^{(1)}  \propto    [N_{\text{corr}} ]^q  \quad \hbox{\ with\ } \quad \frac{2}{3} \le q \le 2 
\label{eq13b}
\end{equation}

We emphasize that as the exponent $q$ cannot be zero, Eq. \ref{eq13b} establishes that \textit{the aforementioned entropic contribution to the cubic 
susceptibility is connected} to $N_{\text{corr}}$. 

\subsubsection{\label{partA2-2} The specific case of RFOT.}

The highest possible value $q=2$ in Eqs. \ref{eq12b}-\ref{eq13b} corresponds to the Adam Gibbs argument $S_c \propto 1/N_{\text{corr}}$. In the original Adam-Gibs argument \cite{Ada65,Bou04,note3}, this comes 
from the assumption that the barrier height $\Delta$ governing relaxation is proportionnal to $N_{\text{corr}}=(\ell /a)^d$. 
This of course overestimates $\Delta$ since it supposes that any relaxation involves a finite fraction of the total number of molecules in the domain. 
This is unlikely when $\ell$ is large since, on general grounds, the energy cost to relax a domain of 
size $\ell$ cannot scale more rapidly than $(\ell/a)^{(d-1)}$, see \cite{note4}. 
 
It does not seem easy to combine this idea that the energy cost scales as $\propto (\ell /a)^\theta$ where $\theta$ \textit{must be} lower than or equal to $(d-1)$ with the result 
 that $\ln(\tau_{\alpha}/\tau_0) \propto 1/(TS_c)$ which, as aforementionned, is obeyed on a vast series of glass forming liquids \cite{Ric98}. The Random First Order Transition 
 theory \cite{RFOT}, proposed thirty years 
 after Adam Gibbs seminal paper, achieves this task, as we briefly recall now. According to RFOT, 
if a domain of size $\ell$ relaxes, the associated free energy cost is given by $\delta F = \Upsilon (\ell / a)^\theta - T S_c (\ell /a)^d$ where the first term corresponds to the 
surface energy cost and the second term is the free energy gain of entropic origin. Relaxation can happen only provided that states are 
available, i.e. for large enough $\ell$. One shows \cite{Bou04} that this happens when $\ell \ge \ell^\star$  where $\ell^{\star}$ is the maximum of $\delta F(\ell)$. 
As the relaxation time $\tau$ is given by $\ln{(\tau/ \tau_0)} = (\ell/a)^\Psi \Delta_0 /(k_B T)$, the typical domain size is $\ell^\star$ because it is the one 
which minimizes $\tau$. One finds:   

\begin{eqnarray}
\left( \frac{\ell^{\star}}{a} \right)^d &=&  \left[ \frac{\theta \Upsilon}{d T S_c} \right]^{\frac{d}{d - \theta}} \label{eq14} \\
\ln{\frac{\tau_{\alpha}}{\tau_0}} &=& \frac{\Delta_0}{k_B T} \left[ \frac{\theta \Upsilon}{d T S_c} \right]^{\frac{\Psi}{d - \theta}} \label{eq15}
\end{eqnarray}

Using this relation and that 
$$ N_{\text{corr}} = \left( \frac{\ell^{\star}}{a} \right)^d $$
one finds eq. (\ref{eq12bis}) for the entropic contribution.  

Within RFOT it has been argued that $\Psi = \theta = d/2$ and that $\Upsilon = \kappa T$ with $\kappa$ a quasi universal constant \cite{RFOT}. This leads to the Adam Gibbs 
relation, i.e. $\ln(\tau_{\alpha} /\tau_0) \propto 1/(TS_c)$, and to a $\chi_{2,1}^{(1)}$ \textit{ which is directly proportional to} $N_{\text{corr}}$.


\subsection{\label{partA3} Box model predictions for $X_3^{(1)}$.}

In the Box model, the supercooled liquid is assumed to be made of independent Dynamical Heterogeneities (DH). Each DH has its own dielectric relaxation time $\tau_{dh}$ and the probability distribution $\cal G$ of the $\tau$'s is choosen so as to recover the linear polarisation of the material. The dynamics of the polarization of a given DH is assumed to be given by a Debye equation, i.e. its time dependent linear polarization $P_{lin,dh}(t)$ is given by:

\begin{equation}
P_{lin,dh}(t) = \frac{\epsilon_0 \Delta \chi_1 E_{ac}}{\sqrt{1+y^2}} \cos(\omega t - \arctan(y))
\label{BM1}
\end{equation}
 where we have set $y = \omega \tau_{dh}$.
 
 The key assumption of the Box model is about the dissipated heat which is supposed to remain confined within each DH during a thermal time equal to the dielectric 
 time $\tau_{dh}$. Because the dissipated power $\cal P$ is quadratic in the field $E_{ac}$, it contains a static term ${\cal P}_0$ and a term ${\cal P}_2$ oscillating at $2 \omega$. The resulting heating of the considered DH contains the two corresponding terms $\delta T_{0,dh}$ and $\delta T_{2,dh}$. 
 
 Because the Box model has been shown to be efficient only for $X_3^{(1)}(f \gg f_{\alpha})$ we shall focus onto $X_3^{(1)}$ and use the fact that $f \gg f_{\alpha}$ amounts to $y \gg 1$ for most of the DH's --this comes from the shape of ${\cal G}(\tau)$ which has a sharp peak around $\tau_{\alpha} = 1/(2\pi f_{\alpha})$. As shown in  Eqs. 4-5 of Ref. \cite{Bru11b}, as soon as $y \gg 1$ one has  $\delta T_{2,dh} \ll \delta T_{0,dh}$ and $X_{3}^{(1)}$ can be computed just by using the values of $\delta T_{0,dh}$ in the various DH's. For any given DH where $y \gg 1$ one gets \cite{Bru11b}:
 
\begin{eqnarray}
P_{3,dh}^{(1)}(t) & \simeq & \frac{\partial P_{lin,dh}(t)}{\partial T} \delta T_0 \label{BM3} \\
\delta T_{0,dh} &=& \frac{\epsilon_0 \Delta \chi_1 E_{ac}^2}{2c_{dh}} \frac{y^2}{1+y^2}  \label{BM2}
\end{eqnarray}
where $c_{dh}$ is the part of  the -volumic- specific heat involved in the DH's and $P_{3,dh}^{(1)}$ is, for a given DH, the term corresponding to the first one of the right hand side 
of Eq. \ref{eq19} in the main text. In complete analogy with  Eqs. \ref{eq19}-\ref{eq8} of the main text, we give here the definition of the dimensionless cubic susceptibility $\vert X_{3,dh}^{(1)}\vert e^{-i \delta_{3,dh}^{(1)}}$ for a given DH, namely:

\begin{equation}
P_{3,dh}^{(1)}(t) = \frac{3}{4} \epsilon_0 \frac{\epsilon_0 \Delta \chi_1^2 a^3}{k_B T} E_{ac}^3 \vert X_{3,dh}^{(1)} \vert \cos(\omega t - \delta_{3,dh}^{(1)})
\label{BM4}
\end{equation}

We just have now to combine Eqs. \ref{BM1}-\ref{BM4} to obtain:
\begin{eqnarray}
\vert X_{3,dh}^{(1)} \vert e^{ - i\delta_{3,dh}^{(1)}} &\simeq& \frac{4}{3} \frac{k_B}{c_{dh}a^3} \vert T \chi_T \vert \frac{y^3 \sqrt{4y^2+(y^2-1)^2}}{(1+y^2)^3} \nonumber \\
\ &\ &  \times e^{- i\arctan\left[ \frac{y^2-1}{2y}\right]}
\label{BM5}
\end{eqnarray}
where we have set $\vert T \chi_T \vert = - \partial \ln(\tau_{dh})/(\partial \ln(T))$ because we assume that the Time-Temperature Superposition property (TTS) holds, i.e. that all the $\tau$'s evolve in temperature as the typical 
relaxation time $\tau_{\alpha}$. Apart from that, in Eq. \ref{BM5} the approximate inequality comes only from the fact that we have neglected the subleading term in $y$ coming from  $\delta T_{2,dh}$ in Eq. \ref{BM3}.
  
We now fully simplify Eq. \ref{BM5} by taking the limit $y \gg 1$:

\begin{equation}
\lim_{y \gg 1} \vert X_{3,dh}^{(1,BM)} \vert e^{- i \delta_{3,dh}^{(1,BM)}} = \frac{4}{3} \frac{k_B}{c_{dh}a^3} \vert T \chi_T \vert \frac{e^{- i \pi/2}}{y}
\label{BM6}
\end{equation} 
where the superscript ``BM'' stands for Box model to make the distinction with the Toy model that we are using hereafter.

The Toy model, introduced in Ref. \cite{Lad12}, starts with the same assumption as the Box model regarding the decomposition of a supercooled liquids into independent DH's of distribution ${\cal G}(\tau)$. The dynamical equation for the polarization in the Toy model is a simple Debye equation for the linear response, but when considering higher order responses, new nonlinear terms arise in the equation. As explained in the main text, these nonlinear terms do not come from heatings at the scale of each DH but from the key assumption of the Toy model, i.e. from the assumption that each amorphously ordered DH has a dipole moment $\propto \sqrt{N_{corr}}$. This yields  generically $X_{2k+1} \propto N_{corr}^k$. By using Eq. A29 of Ref. \cite{Lad12}, one has:

\begin{eqnarray}
 \vert X_{3,dh}^{(1,TM)} \vert e^{- i\delta_{3,dh}^{(1,TM)}}&=&\left[ \frac{3}{5}\right]\frac{N_{corr}}{(1-\delta^2)}\frac{\vert {\cal D}_{3,dh}^{(1)}(y)\vert}{\sqrt{1+y^2}}
  \nonumber \\
\ & & \times e^{i\Psi_{3,dh}^{(1)}(y)-i\arctan(y)} \nonumber \\
 \lim_{y \gg 1} \vert {\cal D}_{3,dh}^{(1)}(y)\vert e^{i\Psi_{3,dh}^{(1)}(y)} &=& \frac{1}{2}
\label{TM1}
\end{eqnarray}
where the superscript ``TM'' stands for Toy model.

By using Eqs. \ref{BM6},\ref{TM1} we obtain:

\begin{eqnarray}
\lim_{y\gg 1}\left[ \frac{\vert X_{3,dh}^{(1,TM)} \vert e^{- i\delta_{3,dh}^{(1,TM)}}}{\vert X_{3,dh}^{(1,BM)} \vert e^{- i \delta_{3,dh}^{(1,BM)}}}\right] &=& \frac{9}{40}\left[\frac{c_{dh}a^3}{k_B}\right]\frac{N_{corr}}{(1-\delta^2)\vert T \chi_T \vert} \nonumber \\
\ &\simeq & 2.2 \frac{N_{corr}}{(1-\delta^2)\vert T \chi_T \vert}
\label{BM7}
\end{eqnarray}
where the last aproximate equality was obtained by using the values for glycerol $c_{dh} \simeq 1.2 \times 10^6 J/($Km$^3)$ and $a^3 \simeq 1.2 \times 10^{-28}$m$^3$. 
We thus have shown that, \textit{provided that $N_{corr}$ is proportional to $\vert T \chi_T\vert$}, the predictions of the Box and Toy model for $X_{3}^{(1)}(f \gg f_{\alpha})$ are 
similar \textit{in phase and in magnitude}. This is important since this corresponds to the observable and to the frequency range where 
the Box model has been able to fit the experimental data.  We end with two remarks:

$\bullet$ In glycerol around $204$K one finds $\vert T \chi_T \vert \simeq 10^2$ and in Ref. \cite{Lad12} a good fit of the measured $X_3^{(1)}$ was obtained within the Toy model with $N_{corr} = 15$ and $\delta = 0.6$. Using these values in Eq. \ref{BM7} yields $X_3^{(3),TM}/X_3^{(3),BM} \simeq 0.5$, i.e. a value that is two times smaller than expected. This shows that the limit $y \gg 1$ is not precise enough to give the exact prefactors. Similarly the phase of the measured $X_{3}^{(1)}(f \gg f_{\alpha})$ is not $- \pi/2$ but a factor of 2 smaller. We note furthermore that the exact value of $c_{dh}$ choosen to fit the $X_{3}^{(1)}$ data within the Box model depends on the material and that an adjustable factor -between $0.5$ and $1$- has been used in Ref. \cite{Wan07}. This adjustable factor is within the range of numercial uncertainty produced by our method using the limit $y \gg 1$.

$\bullet$ Also applying this reasoning to $X_3^{(3)}$ does not yield a corresponding result since 
$\vert X_3^{(3),TM}/X_3^{(3),BM}\vert \propto y$, i.e. the two models never yield the same functional dependence on $y$ for the third harmonics cubic susceptibility.


\begin{thebibliography}{99}
\bibitem{note6} As shown in the Appendix, within the standard RFOT where $\Psi \equiv \theta \equiv d/2$, the barrier height is proportional to $(\ell^\star / a)^{d/2}$, yielding 
a much slower dependence than the $(\ell^\star/a)^d$ dependence assumed in the Adam Gibbs theory. One could naively think that such a large difference can be easily 
discriminated experimentally by comparing the $N_{\text{corr}}(T)$ -drawn from $\chi_3(T)$ measurements- and $\chi_T \equiv \vert \partial \ln (\tau_{\alpha}) / (\partial T) \vert$. 
This is not the case since by adopting Adam Gibbs theory, one predicts $N_{\text{corr}} \propto T^2 \chi_T = \vert \partial \ln (\tau_{\alpha}) / (\partial [1/T]) \vert$; 
but this is also the case if the usual form $S_c(T) = S_{\infty} (1-T_K/T)$ is inserted in RFOT, i.e. in Eq. \ref{eq15} with the ``recommended'' values $\Psi = \theta = d/2$. Distinguishing between Adam Gibbs and RFOT thus requires 
more than the study of $N_{\text{corr}}(T)$. This is not so surprising since both theories agree on a key point, namely the existence of a domain length $\ell^\star$ which grows
 upon cooling.
\bibitem{note9} It would be worth to develop further the MCT analysis of \cite{Tar10} to obtain detailed analytical predictions on the phases.
\bibitem{note12} The growth of $\chi_3$ and $\chi_5$ can be induced by purely dynamical correlation, as it is the case in MCT theory and in the MCT regime. 
However, in the activated dynamics regime, the only theory that can explain a growth of $\chi_3$ and $\chi_5$ are the ones in which 
dynamical correlations are driven by static (``point-to-set'') correlations.
\bibitem{values} The minimal value of $q$ is obtained with $\Psi =0$ and $\theta = d-1$, while the maximal value of $q$ comes from $\Psi = d$ and $\theta = d/2$. For $d=3$ this yields $1/3 \le q \le 3/2$.
\bibitem{note10}The only sound physical interpretation we can find is that this heat is related to the part of the entropy, the configurational entropy, that is gained once the system has time to relax and sample all possible metastable states. This is naturally encoded in RFOT and associated to the growth of $N_{\text{corr}}$.
\bibitem{note3} The Adam-Gibbs theory assumes that the barrier height $\Delta$ for a system to relax is proportionnal to $N_{\text{corr}}$ and given by $\Delta_0 (\ell/a)^d$, 
with $\Delta_0$ a constant. By using $\ln(\tau_{\alpha}/\tau_0) =  \Delta /(k_B T) =  (\ell/a)^d \Delta_0/(k_B T)$, one understands that the relaxation processes will be 
overwhelmingly dominated by the fastest ones, i.e. by those corresponding to the smallest possible value of $\ell$. This value is obtained by replacing the inequality by an equality 
in Eq. \ref{eq10}, which yields $(\ell / a)^d =k_B \ln{2} /S_c$ and finally $\ln(\tau_{\alpha}/\tau_0 = \ln{(2)} \Delta_0 /(T S_c)$, which is similar to Eq. \ref{eq2}, 
up to a redefinition of $\Delta_0$ to absorb the factor $\ln 2$.
\bibitem{note4} This can be understood by considering an Ising model -of coupling constant $J$- with a domain 
containing $(\ell/a)^d$ up spins in a sea of down spins: the energy cost to split this domain in two is $\propto J (\ell /a)^{(d-1)}$, then the cost of spliting the two 
new domains of size $\ell /2$ is $\propto J ((\ell/2) /a)^{(d-1)}$, and so on until all the spins have been flipped down. Summing up all these energy costs leads to a 
full energy cost of domain flipping which goes as $(\ell /a)^{(d-1)}$.
\end{thebibliography}
\end{document}